# Polytypes of long-period stacking structures synchronized with chemical order in a dilute Mg-Zn-Y alloy

E. Abe [a,*], A. Ono [a], T. Itoi [b], M. Yamasaki [c] and Y. Kawamura [c]

[a] Department of Materials Science & Engineering, University of Tokyo, Tokyo, 113-8656, Japan
[b] Department of Mechanics Engineering, Chiba University, Chiba 263-8522, Japan
[c] Department of Materials Science, Kumamoto University, Kumamoto 860-8555, Japan

A series of structural polytypes formed in an Mg-1at.%Zn-2at.%Y alloy has been identified, which are reasonably viewed as long-period stacking derivatives of the *hcp* Mg structure with alternate AB stacking of the close-packed atomic layers. Atomic-resolution Z-contrast imaging clearly revealed that the structures are long-period chemical-ordered as well as stacking-ordered; unique chemical order along the stacking direction occurs as being synchronized with a local faulted stacking of AB'C'A, where B' and C' layers are commonly enriched by Zn/Y atoms.

*Corresponding author : abe@material.t.u-tokyo.ac.jp

Recently, dilute magnesium alloys containing a few atomic percent of Zn and Y (or rare-earth (RE) elements) have attracted increasing attention because of their excellent mechanical properties. The first example was demonstrated for a rapid-solidification (RS) processed Mg-1at.%Zn-2at.%Y alloy (denoted as $Mg_{97}Zn_1Y_2$ hereafter), which realizes a maximum tensile yield strength ~600MPa with elongation ~5 % at room temperature [1]. This finding led to a further exploration of various Mg-Zn-RE alloys of similar compositions, and it has been found that many of these alloys are also able to reveal the strength higher than ~350MPa by simply applying a conventional hot-extrusion process for the as-cast ingots [2]. The remarkable microstructural feature common for all the above Mg-Zn-Y(RE) alloys is formation of a novel type of long-period structures [3-6], which are believed to play a critical role to realize the superior mechanical properties [7]. These novel structures are long-period stacking-ordered variants of a hexagonal-close-packed (*hcp*) structure of a Mg crystal, and accordingly rhombohedral (R) and hexagonal (H) Bravais lattices alternatively appear depending on the stacking period of the close-packed atomic layers. So far, four polytypes, 10H, 18R, 14H, and 24R have been reported for the Mg-Zn-Y alloys [8]. Here, we should mention that the 6H-type structure firstly identified [3, 4] was perhaps a misleading of the 18R-type structure; nanometer-scale crystalline grains were heavily faulted in the RS-processed $Mg_{97}Zn_1Y_2$ alloy, and hence the relevant diffraction peaks are strongly streaked. Although a precise assignment of the stacking sequence was difficult, Z-contrast scanning transmission electron microscope (STEM) observation clearly showed up a veiled additional order along the stacking direction [4]. Namely, the long-period Mg-Zn-Y structure is found to be chemical-ordered as well as stacking-ordered, as displayed by a solute element enrichment at the particular layers that are relevant to faulting of the original 2H-type stacking (i.e., *hcp* structure);

a unique long-period stacking/order (LPSO) structure (this wording follows the sense that, in metallurgy, the "order" specifically means an ordered arrangement of different atom species (i.e., chemical order in general) on the fixed atomic sites). The following Z-contrast STEM observations clearly identified the plausible stacking sequences as well as the significant Zn/Y enrichment at the fault-related layers, which form the definite structural unit common for the 18R-type and 14H-type LPSO structures [9-11].

In the present paper, we systematically investigate all the structural polytypes presently available, 10H, 18R, 14H, and 24R formed in the $Mg_{97}Zn_1Y_2$ alloy, by focusing particularly on the chemical order feature along the long-period stacking direction. As described in the followings, Z-contrast STEM imaging have successfully identified that all these polytypes are constructed by the common structural unit [9-11] composed of local fault-related layers that are essentially enriched by the Zn/Y atoms. Consequently, the present results firmly identify a unique series of polytypes - the synchronized LPSO structures, for which the chemical order occurs so as to synchronize with the relevant stacking order.

A master alloy ingot of $Mg_{97}Zn_1Y_2$ was prepared by high-frequency induction melting of pure metals in an argon atmosphere. Rapidly-solidified ribbons were prepared by a single-roller melt-spinning method with wheel rotating speed of approximately 42 m/s. RS ribbons were annealed at several temperatures to obtain the structural polytypes (see ref. 12 for details), and the annealed specimens were thinned by argon ion milling. Z-contrast STEM observations were performed by a conventional 200 kV STEM (JEM-2010F), which provides a minimum probe of approximately ~0.15 nm with a convergence semi-angle ~10 mrad. The annular-detector was set at the scattering-angle range 70-150mrad being sufficiently high to reveal a chemical sensitive Z-contrast.

Figures 1 (a)-(e) show the electron diffraction (ED) patterns and the corresponding Z-contrast STEM images of the Mg-Zn-Y long-period structures, together with those of the Mg *hcp*-structure as a reference. The *hcp* structure is constructed by stacking the close-packed layers with a two-layer periodicity (Fig. 1(a)), and its long-period stacking derivatives are generated by introducing the stacking faults periodically into the original *hcp* crystal. For the Mg-Zn-Y polytypes, it is found that the stacking faults are introduced at every 5-, 6-, 7- and 8-layer, forming the alternate H-R series of 10H, 18R, 14H and 24R, respectively, as represented by the atomic models inserted in the images (Figs. 1(b)-(e)). Note that the stacking sequences of the present 14H and 18R models are identical with that reported in ref. 6 among the several sequences proposed in early studies [5, 6, 8]. These 14H/18R stacking models, which have been also confirmed by the following Z-contrast observations [9-11], reasonably and systematically explain a series of the Mg-Zn-Y polytype structures. In the Z-contrast images, significant bright contrast representing Zn/Y enrichment occurs at the stacking fault and its neighbor layers, as being common for all the long-period prolytypes. Consequently, the correlation length of the chemical order always appears to be coincident with that of the relevant stacking order. On this basis, it is evident that all the Mg-Zn-Y polytypes definitely belong to the LPSO structure with a synchronized stacking/chemical order, for which the 00$l$ systematic reflections intrinsically appear [4] as shown by inset in the images.

Figure 2 shows the model structures of the 10H, 18R, 14H and 24R with their cyclic stacking represented by A-B-C sequences. We find that a series of these LPSO structures are systematically described by the common structural unit composed of local AB'C'A stacking [9-11] (shown upper-left in Fig. 2), where C represents the fault layer with respect to the original AB stacking, and B' and C' denote the layers enriched by Zn and Y. Occurrence of such Zn/Y enrichment at the two distinct layers can be understood by considering their local environments; see the *h* and *c* notations (Jagodzinski notation) attached to the ABC stacking columns in Fig. 2. Here, *h* means the local *hcp* environment and is hence attributed to the layer sandwiched by the same layers; A<u>B</u>A where B layer is assigned as *h*. Similarly, *c* means the local face-centered-cubic (*fcc*) environment, and therefore the *c* is relevant to the layer sandwiched by the different layers; A<u>B</u>C where B layer is assigned as *c*. With this assignment, it turns out that the Zn/Y enrichment essentially occurs at the *c* layers of a local *fcc* environment; namely, the AB'C'A stacking unit [9-11] is represented as *hcch* (note that the atomic sites within these *c* layers can be equivalent for a certain space group; e.g., $R\bar{3}m$ for the R-type structures).

Even though the Z-contrast at the *c* layers in the LPSO structures appear to be remarkably strong (Fig. 1), the atomic sites within these layers are not fully occupied by Zn and Y atoms [4, 13]; the compositions of the Zn/Y enriched layers were previously estimated to be approximately ~5at.% Zn and ~10at.% Y ($Mg_{85}Zn_5Y_{10}$) [4]. It is noteworthy here that the compositions of the *c* layer seem to be almost identical for all the present LPSO polytypes; see Fig. 3. Z-contrast images were obtained from the region where different types of the LPSO structures coexist in the lamellar packets [12] (Fig. 3(a)), and therefore the continuous intensity profiles can be traced across the different period structures. As shown in Fig. 3(b)-(d), the intensity at the Zn/Y-enriched *c* layers appears to be almost identical for all the LPSO structures. This means that the compositions of the LPSO phases could be slightly different depending on their periodic-length. Assuming that the *c*-layer composition is $Mg_{85}Zn_5Y_{10}$ and all the *h* layers are purely occupied by Mg atoms, the phase compositions are estimated to be $Mg_{94.0}(Zn,Y)_{6.0}$, $Mg_{95.0}(Zn,Y)_{5.0}$, $Mg_{95.7}(Zn,Y)_{4.3}$ and $Mg_{96.3}(Zn,Y)_{3.7}$ for the 10H-, 18R-, 14H- and 24R-type structures, respectively, as indicated at the bottom in Fig. 2. On the basis of this fact, we shall identify the present LPSO variations as polytypes with a broad definition, since, in a strict sense, the polytypes (polymorphism) should refer a composition-invariant series.

In contrast to the recent observations [11], the appearances of superlattice reflections in the close-packed planes are extremely weak for the present LPSO phases, as exemplified by the diffraction patterns of the H-type LPSO phase; see Fig. 4 (note that the incidence normal to the close-packed planes can be coincident with a low-index crystallographic zone axis only for the hexagonal lattice, and not for the rhombohedral lattice). This significantly less-ordered state, which was also confirmed for the similar RS-processed $Mg_{97}Zn_1Y_2$ alloy [14], may be related to the compositions of the Zn/Y-enriched layers. That is, distinct superlattice reflections appears for the LPSO phase whose relevant layer is sufficiently enriched by Zn/Y, e.g., ~$Mg_{50}Zn_{25}Y_{25}$ [11], which is remarkably higher than that of the present LPSO phases ($Mg_{85}Zn_5Y_{10}$) in the RS-processed $Mg_{97}Zn_1Y_2$ alloy. It is also noteworthy that formation behaviors of the LPSO phases were found to be considerably different between the RS-ribbon and bulk-ingot specimen [12], perhaps being due to different impurity contents such as oxygen. As preliminary models, we presently assume the random distribution of Zn and Y [4] within the $c$ layers for calculating structure factors, reproducing fairly well the observed diffraction patterns, as shown in Fig. 2. On the basis of further precise/quantitative analysis, details of Zn/Y chemical order and the compositions of the relevant layers will be systematically investigated for the LPSO phases with different Zn/Y contents in the forthcoming paper [15].

Finally, we briefly describe the characteristics of the present LPSO structures by comparisons with some alloy structures. Solute-atom enrichment at the particular atomic planes is a well-known phenomenon such as G.P. zones in dilute Al alloys. However, there are no long-range correlations developed between the solute-enriched layers; e.g., Cu-clustering occurs randomly at the {100} planes of the *fcc* matrix and hence gives rise just streaks in the diffraction pattern. The long-period series reported for the $Ni_3$(Ti, Nb) alloys [16] represents the stacking variants of the original $D0_{19}$-type structure (chemically ordered version of the *hcp*), for which the every stacking layer is composed of all the constituent elements with chemical ordered arrangements. Therefore, these can be viewed as the LPSO-type but being without synchronization of stacking/chemical order; i.e., no superlattice peaks appear along the $00l$ systematic reflections (compositions are identical for all the layers). Long-period Mg-In structures may also belong to this non-synchronized LPSO-type [17]. Perhaps there may be more example structural series similar to the present LPSO polytypes, we nevertheless emphasize a synchronized occurrence of stacking/chemical order for the present Mg-Zn-Y LPSO phases. Because of this unique order with a confined Zn/Y distribution, the LPSO phases are able to form with less Zn/Y contents (Fig. 2), enabling to gain a considerable amount of their volume even in the dilute Mg-Zn-Y(RE) alloys. This is important for designing the (LPSO + α-Mg) two-phase practical alloys [2], whose mechanical properties can be tuned by a phase-volume ratio along with the concept of composite materials [7]. Full description of the LPSO structure, including space-group symmetry and the precise Zn/Y arrangements as well as Mg atomic sites, will be described in the forthcoming paper [15].

**Acknowledgments**

We are grateful to Professors Hagihara, Ohtani, Higashida for stimulating discussions. This work is supported by a Grant-in-Aid for Scientific Research on Priority Areas "Atomic Scale Modification'' from the Ministry of Education, Culture, Sports, Science and Technology of Japan (MEXT), and the Kumamoto Prefecture CREATE Project from JST.

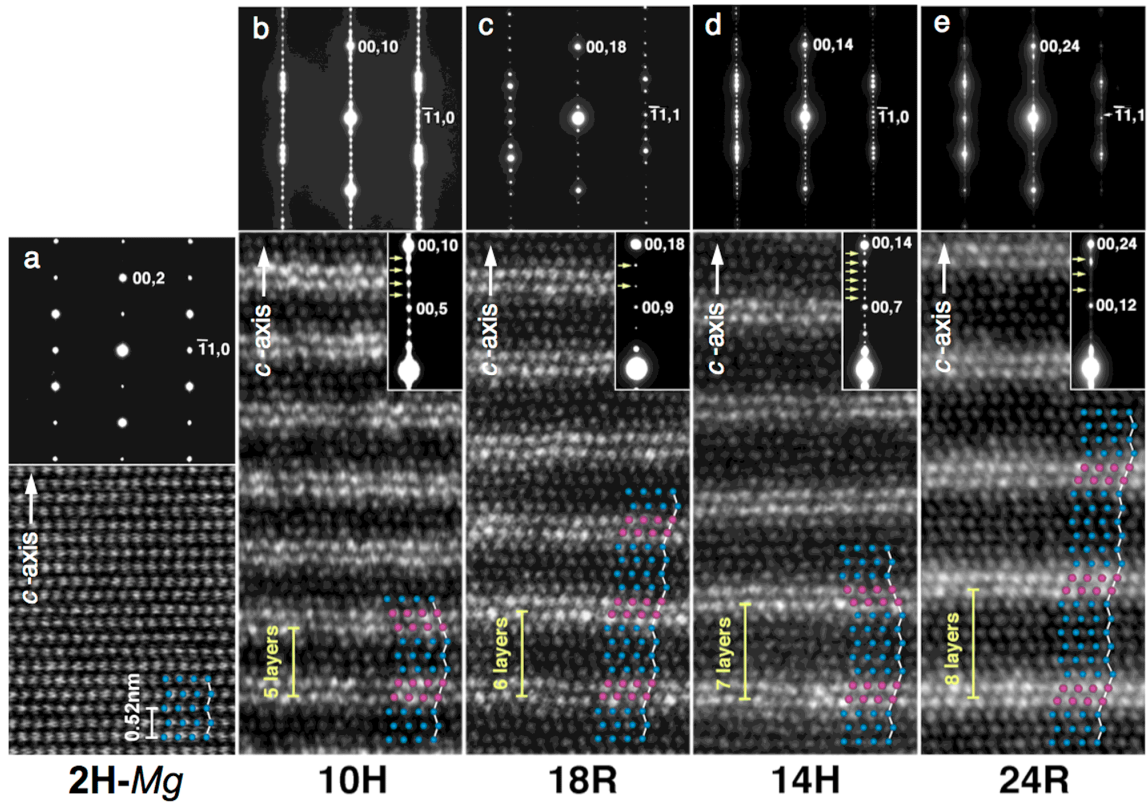

**Figure 1.** Electron diffraction patterns and Z-contrast STEM images of (a) *hcp*-Mg and long-period structures of (b) 10H-type, (c) 18R-type, (d) 14H-type and (e) 24R-type. These images are taken with the incident electron beam along the directions relevant to [11,0] axis of the *hcp* host crystal, revealing the definite stacking sequences of the polytypes. These long-period structures were formed in the RS $Mg_{97}Zn_1Y_2$ alloy annealed (b) at 573K for 1 h and (c) - (d) at 673K for 48 h [12].

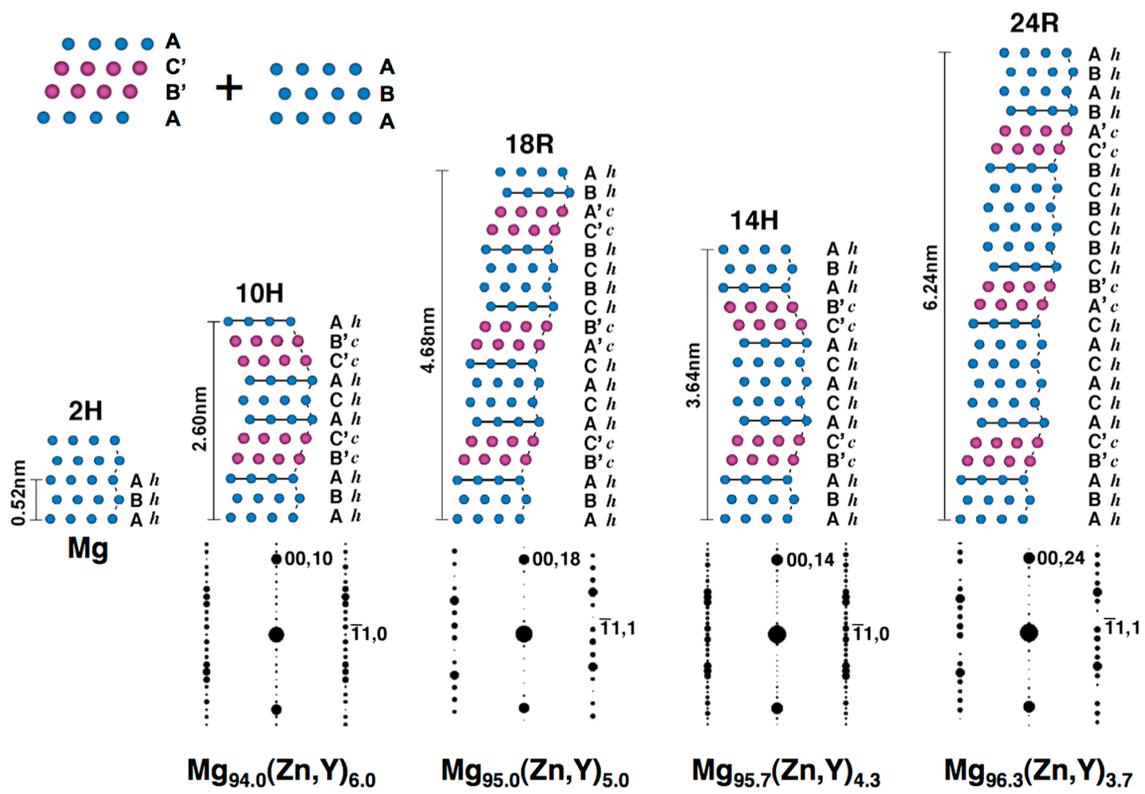

**Figure 2.** Structure models of the Mg-Zn-Y LPSO polytype structures. Blue and red circles represent the Mg site and Zn/Y occupation site, respectively. Details are described in the text.

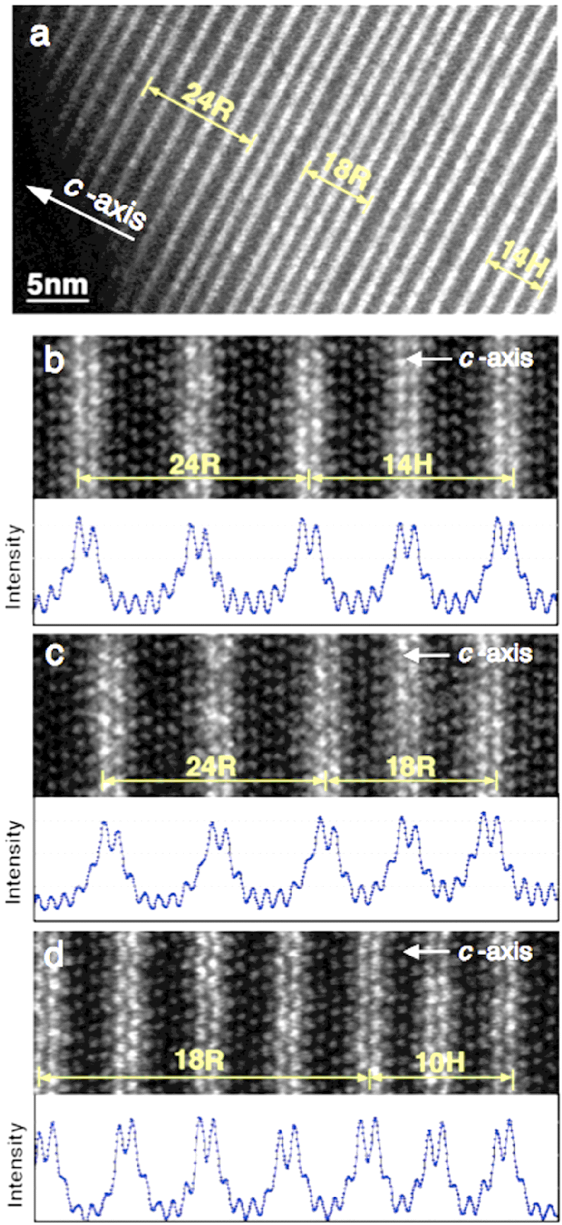

**Figure 3.** (a) Z-contrast STEM image of the LPSO multiple-phase region, taken from the RS $Mg_{97}Zn_1Y_2$ alloy annealed at 673K for 48 h. (b) – (d) Continuous Z-contrast intensity profiles across the different LPSO structures, where each peak represents the layer-intensity integrated over the relevant close-packed layers within the image area; (b) 24R-14H, (c) 24R – 18R and (d) 18R – 10H. These images are taken along the directions relevant to [11,0] axis of the *hcp* host crystal.

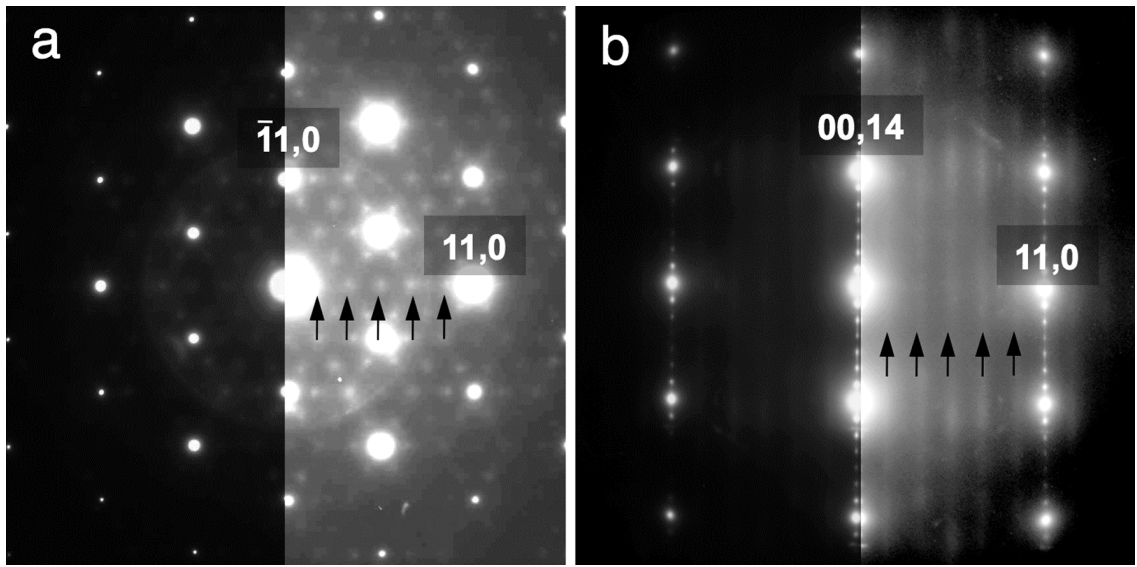

**Figure 4.** Electron diffraction patterns of the H-type LPSO phase in the RS $Mg_{97}Zn_1Y_2$ alloy annealed at 673K for 48 h, taken with the incident electron beam along (a) [00,1] and (b) [1$\bar{1}$,0] axes. For each pattern, the intensity is shown by linear-scale and logarithmic-scale at left-half and right-half sides, respectively, in order to show up the extremely weak superlattice diffuse peaks/streaks indicated by arrows.